\documentclass[superscriptaddress,twocolumn,prb,longbibliography]{revtex4-2}
\usepackage{graphicx}
\usepackage{amsmath}
\usepackage{amsthm}
\usepackage{amssymb}
\usepackage{amsfonts}
\usepackage{siunitx}

\usepackage{blindtext}
\usepackage{comment}
\usepackage{xcolor}
\usepackage{xr}
\usepackage{lineno}
\usepackage[colorlinks]{hyperref}

\newcommand{\unit}[1]{\,\mathrm{#1}}

\newcommand{\odr}{\omega_{\rm dr}}

\begin{document}

\title{{\bf A chemical nano-reactor  based on a levitated nanoparticle in vacuum}}

\author{F. Ricci}
\affiliation{ICFO-Institut de Ciencies Fotoniques, The Barcelona Institute of Science and Technology, 08860 Castelldefels (Barcelona), Spain}

\author{M. T. Cuairan}
\affiliation{ICFO-Institut de Ciencies Fotoniques, The Barcelona Institute of Science and Technology, 08860 Castelldefels (Barcelona), Spain}
\affiliation{ETH Z\"urich, Nanophotonic Systems Laboratory, 8092 Z\"urich, Switzerland}

\author{A. W. Schell}
\affiliation{ICFO-Institut de Ciencies Fotoniques, The Barcelona Institute of Science and Technology, 08860 Castelldefels (Barcelona), Spain}
\affiliation{Institut f\"ur Festk\"orperphysik, Leibniz Universität Hannover, 30167 Hannover, Germany}
\affiliation{Physikalisch-Technische Bundesanstalt, 38116 Braunschweig, Germany}

\author{E. Hebestreit}
\affiliation{ETH Z\"urich, Photonics Laboratory, 8093 Z\"urich, Switzerland}

\author{R. A. Rica}\thanks{Corresponding author: rul@ugr.es}
\affiliation{Universidad de Granada, Nanoparticles Trapping Laboratory and Research Unit Modeling Nature (MNat), 18071, Granada, Spain}
\affiliation{Universidad de Granada, Department of Applied Physics, 18071, Granada, Spain}

\author{N. Meyer}\thanks{Corresponding author: n.meyer@ethz.ch}
\affiliation{ICFO-Institut de Ciencies Fotoniques, The Barcelona Institute of Science and Technology, 08860 Castelldefels (Barcelona), Spain}
\affiliation{ETH Z\"urich, Nanophotonic Systems Laboratory, 8092 Z\"urich, Switzerland}


\author{R. Quidant}
\affiliation{ICFO-Institut de Ciencies Fotoniques, The Barcelona Institute of Science and Technology, 08860 Castelldefels (Barcelona), Spain}
\affiliation{ETH Z\"urich, Nanophotonic Systems Laboratory, 8092 Z\"urich, Switzerland}
\affiliation{ICREA-Instituci\'o Catalana de Recerca i Estudis Avan\c{c}ats, 08010 Barcelona, Spain}

\begin{abstract}
\noindent 
A single levitated nanoparticle is used as a nano-reactor for studying surface chemistry at the nanoscale. Optical levitation under controlled pressure, surrounding gas composition, and humidity provides extreme control over the nanoparticle, including dynamics, charge, and surface chemistry. Using a single nanoparticle avoids ensemble averages and allows to study how the presence of silanol groups at its surface affects the adsorption and desorption of water from the background gas with unprecedented real time, spatial, and temporal resolution. Here, we demonstrate the unique potential of this versatile platform by studying the Zhuravlev model in silica particles. In contrast to standard methods, our system allowed the first observation of an abrupt and irreversible change in scattering cross section, mass, and mechanical eigenfrequency during the dehydroxylation process, indicating changes in density, refractive index and volume. 
\end{abstract}


\maketitle

\paragraph*{Introduction.}

Surface properties of nanoscale systems govern much of their behavior due to the increased surface-to-volume ratio. Therefore, the broad range of applications based on nanoparticles (NPs)\cite{Cancer,Water} requires surface adaptation in order to develop particular properties \cite{Williams,albanese2012effect}. These tailored NPs are characterized by a large number of different techniques, providing detailed information about matter at the nanoscale \cite{olson2015optical, yurt2012single}. However, most of these techniques average over ensembles of particles, leading to an unavoidable loss of single particle information. Alternatively, electron or dark-field microscopy \cite{Darkfield, Azubel909electron} target individual NPs with highest resolution, but samples must be deposited on a substrate. The substrate itself and neighboring NPs strongly affect the NPs properties. \\
The levitation of individual particles with optical tweezers or Paul traps has become a well established approach to characterize single microparticles in controlled conditions and evades both ensemble averages and disturbance from substrates, therefore enabling fundamental research in aerosol science \cite{krieger2012exploring,gong2018optical} and nanochemistry \cite{howder2015thermally,howder2015single,david2016ultraviolet,hoffmann2020electronic}. In the case of aerosol science, most studies have been performed with single microparticles with diameters larger than $2\unit{\mu m}$. Less is known about the behavior of single particles in the so-called accumulation mode (diameter $0.1-2\unit{\mu m}$), which typically requires ensemble averages to provide meaningful data \cite{cotterell2014measurements}. Indeed, particles in the atmosphere belong predominantly to the accumulation mode and therefore are paramount for its behavior \cite{kaufman2002satellite,reid2018viscosity, tirella2018extending}. Working with even smaller particles (diameter $< 100\unit{nm}$) would also be desirable, since most of the nanoparticles in the accumulation mode that are present in the atmosphere are formed from nucleation of nanoparticles in the range $1-30 \unit{nm}$. Therefore, the properties of these small particles and processes in which they can be involved are of interest since they determine the presence of accumulation-mode particles in the atmosphere \cite{winkler2008heterogeneous}, but their optical levitation and control still remains challenging \cite{millen2020optomechanics,gonzalez2021levitodynamics}.

Single NPs are typically characterized in Paul traps \cite{howder2015single,david2016ultraviolet,hoffmann2020electronic, schell2017flying, kuhlicke2014nitrogen}, while the additional use of optical tweezers enables an increased coupling of the particle with the probing laser beam \cite{conangla2020extending}. In fact, the diameter $d$ of the NP strongly affects the scattering cross section $\sigma_{scat}$, which scales as $d^6$, making small particles hard to detect. 
Moreover, the increased Brownian motion of small NPs prevented their stable trapping with optical tweezers until recently. The young field of levitodynamics has enhanced the manipulation capabilities based on optical tweezers, and now permits stable trapping of NPs in a fully controlled environment in high vacuum \cite{Gieseler2012, Jain2016Direct, Ricci2019Accurate, doi:10.1063/1.5109645,PhysRevLett.121.033602,millen2020optomechanics,gonzalez2021levitodynamics}. As a consequence, levitated NPs have found applications in force and inertial sensing \cite{Hebestreit2018Sensing, Hempston2017Force, Geraci2010}, and have also been proposed to shed light onto the transition between the classical and quantum world \cite{Romero-Isart2011, Bateman2014}. Recent advances in the field allow for contact-less manipulation of the bulk temperature \cite{Hebestreit2018Measuring,coppock2021high} and the NP's charge with single elementary charge accuracy \cite{Frimmer2017Controlling}.

Here, we use the newly achieved level of control to study complex transformations of single NPs involving fast and simultaneous changes in mass $m$, polarizability $\alpha$, 
charge $n_q$, and surface composition. In particular, we investigate how the mass of a silica NP changes due to water uptake (release) from (to) the background gas, which we believe depends strongly on the presence of silanol groups at the surface of the NP \cite{Zhuravlev20110TheSurface}. By increasing the NP's bulk temperature $T_{\text{bulk}}$ to several times the room temperature, we show that the silica NP undergoes an irreversible sudden change in mass, polarisability, charge, and surface composition. We conclude that the high local temperature forces the abrupt desorption of all adsorbed water and the resulting hydrophobic character of the NP hints to a dehydroxylation of the silica surface \cite{Zhuravlev20110TheSurface}. Accurate monitoring of the entire process is made possible by single charge control, and time and mass resolution in millisecond and femtogramm  range, respectively. We anticipate that the demonstrated capabilities based on optically levitated NPs will serve as an enabling technology in the future of nanochemistry.\\

\begin{figure}[h]
	\includegraphics[width=\linewidth]{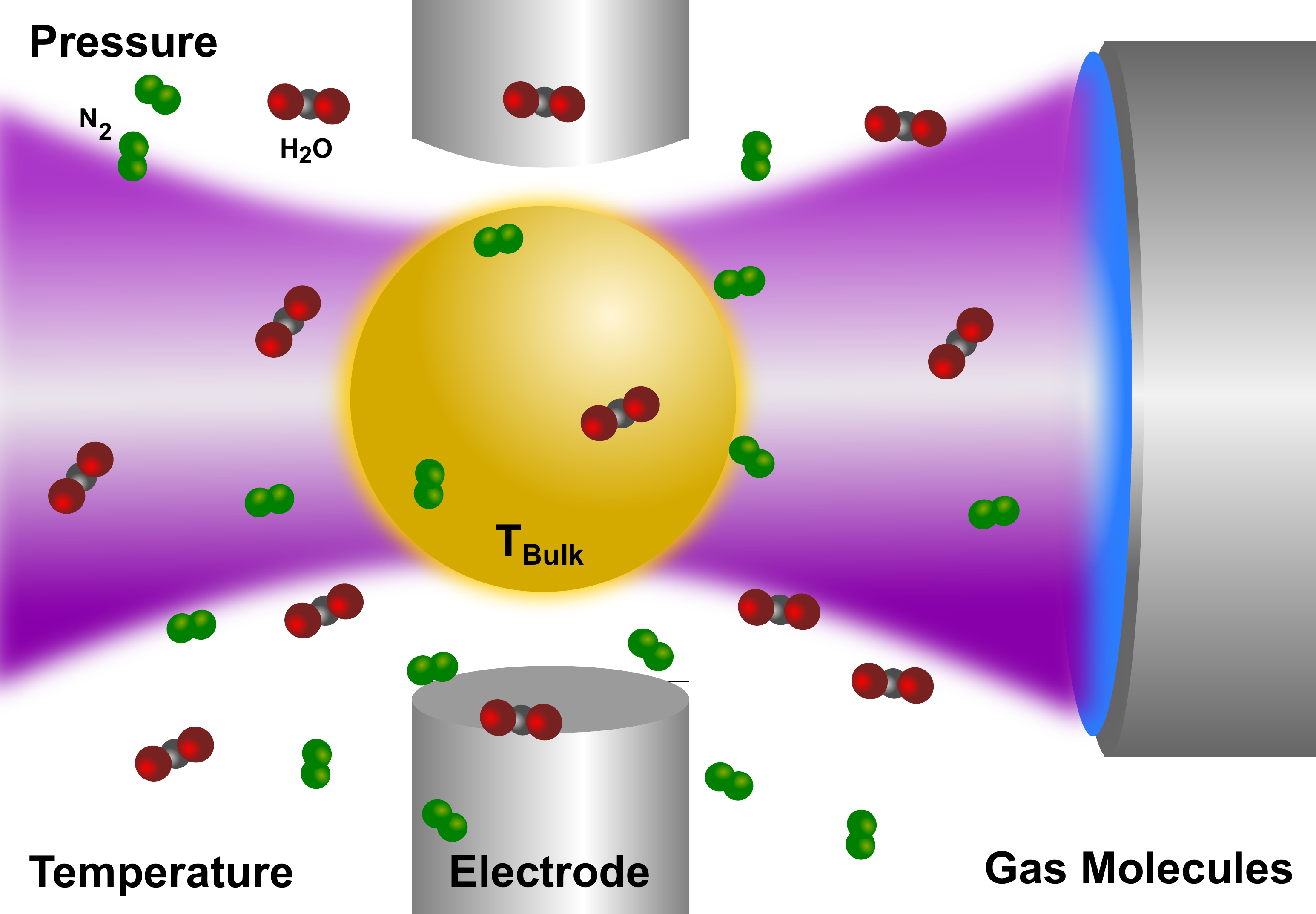}
	\caption{\footnotesize \textbf{Experimental setup} (\textbf{a}) Optical tweezers ($\lambda = 1064\unit{nm}$, NA = 0.8, $P=75\unit{mW}$) levitate a silica NP ($d= 143\unit{nm}$ or 235$\unit{nm}$) between two electrodes that apply an oscillating Coulomb force $\pmb {F_{el}}$ to the NP. We choose the background gas to be either dry  nitrogen (N$_2$) or clean air, in order to control the surrounding humidity level.  The gas pressure $p$ and therefore the bulk temperature $T_{\text{bulk}}$ can also be controlled (see main text). The NP's charge $n_q \times q_e$ is varied by creating a polarised plasma field inside the vacuum chamber (not shown, see SI \ref{app:ChargeControl}).}
	\label{fig:01_ExpSetUp}
\end{figure}

\paragraph*{Experimental platform for chemistry at the nanoscale.}
\noindent We have developed a platform to perform fundamental studies of surface chemistry on a single levitated NP. A schematic representation of our experimental platform is displayed in Fig.\ref{fig:01_ExpSetUp}. A single NP is trapped inside a vacuum chamber by optical tweezers with wavelength $\lambda = 1064\unit{nm}$, power $P\simeq 75\unit{mW}$, and numerical aperture $N\!A=0.8$. The optical gradient force confines the dielectric NP to the focus and thus creates a conservative harmonic trap. We use single amorphous St\"ober silica NPs of density $\rho =2200\unit{kg/m^3}$ and diameter of either $d = 2R = 143\pm 4\unit{nm}$ or 235$\pm 11\unit{nm}$~\cite{microparticles2018Data, Stober1968Controlled}, that are surrounded by a background gas of choice at controllable pressure $p$. In order to vary the water content inside the chamber, we vented with either dry nitrogen N$_2$ or clean air.
The bulk temperature $T_{\text{bulk}}$ of the NP is governed by the balance of absorbed power $P_{\text{abs}}$ and emitted power $P_{\text{emit}}$.
The variable pressure $p$ allows control of $P_{\text{emit}}$ through the number of background gas collisions, while $P_{\text{abs}}$ is constant at constant trapping laser power $P$  (see SI \ref{app:PressTemp}). Notice that, in the range $p\geq 0.1\unit{mbar}$, black-body radiation (absorption  and emission) can be neglected for the laser intensities used. By this method, we control $T_{\text{bulk}}$ over a wide range ($\approx 300-1000\unit{K}$). Alternatively, pressure-independent control techniques of $T_{\text{bulk}}$  via laser absorption \cite{Hebestreit2018Measuring,coppock2021high} or laser cooling \cite{Rahman2017} are readily available. 

We gain information about the NP's polarizability $\alpha  = 4\pi\varepsilon_0 R^3 (n_r^2 -1)/(n_r^2 +2)$, and consequently also about changes in volume $V=\frac{4}{3}\pi R^3$ or refractive index $n_{r}=1.46$, by observing the brightness of the scattered light $\propto |\alpha|^2$ on a CCD camera (see SI \ref{app:Brigthness}). Moreover, any trapped NP carries initially a few elementary charges ($n_q^{\rm in}\simeq5-10$), likely acquired due to the triboelectric effect during the nebulization process used to load the trap (see SI \ref{app:aerosol}). We control the NP's charge by generating a corona discharge inside the vacuum chamber (not shown in Fig.\ref{fig:01_ExpSetUp}, see SI \ref{app:ChargeControl}). The plasma leads to a charge flux that adds elementary charges to the NP's surface and therefore induces a discrete charge change $\Delta n_q \times q_e$, where $n_q$ is the number of elementary charges $q_e = 1.6\times 10^{-19} \unit{C}$ \cite{Frimmer2017Controlling, Ricci2019Accurate}. Finally, the NP position $(x,y,z)$ and its eigenfrequencies $\Omega_{x,y,z}$  are continuously measured (only in the underdamped regime $p \le 50\unit{mbar}$) by interfering the scattered light with a reference beam using a balanced split detection scheme~\cite{Ricci2017Optically, Gieseler2013Thermal}. 

The NP's dynamic response to an oscillating electric force $\pmb{F_{el}}(t)$ gives an estimation of the mass $m$ \cite{Ricci2019Accurate} (see SI \ref{app:Mass}) and charge $n_q$ of the NP \citep{Frimmer2017Controlling} (see SI \ref{app:ChargeControl}). $\pmb{F_{el}}(t)$ is generated by applying an electric field $E(t)=E_0\cos( \odr t )$ to two electrodes, situated above and below the NP (see Fig. \ref{fig:01_ExpSetUp}). In the experiments described in this work, we charge the NP to $n_q = 5-10\times q_e$ to obtain the mass at $p = 50\unit{mbar}$ (see SI \ref{app:Mass}) and we neutralize the NP ($n_q=0$) before lowering the pressure in each cycle. Sudden changes in eigenfrequencies observed at low pressure reveal changes of the NP density and refractive index, since $\Omega_i \propto \sqrt{\frac{n_r-1}{n_r+2}\frac{1}{\rho}}$. \\ 

\paragraph*{Surface chemistry of silica.}
As a proof of principle of the versatile capabilities of our platform, we investigate the surface chemistry of silica. 
The chemistry of silica is important, not just from a fundamental point of view, but also for the large number of applications based on silica nanoparticles \cite{bergna2005colloidal}, including biomedicine \cite{li2012mesoporous,wang2015mesoporous},
food industry \cite{higashisaka2015applications,VIDEIRAQUINTELA2021106318}, cosmetics \cite{fytianos2020nanomaterials}, and materials \cite{VERMA2021,LAMYMENDES2021122815}
. The optical, chemical, and mechanical properties of silica largely depend on its surface chemistry \cite{Cerveny}. 
For amorphous silica, these properties are typically understood in terms of the Zhuravlev model \cite{Zhuravlev1987, Zhuravlev20110TheSurface}. In a nutshell, the Zhuravlev  model describes how dehydration, dehydroxylation, and rehydroxiylation occur upon thermal treatments of the silica surface. Silanol groups are formed on the NP surface during the silica synthesis process. If the concentration of silanols is large enough, they build a hydrophilic layer on the surface that allows for water adsorption. The removal of these hydroxyl groups from the surface leads to a decrease in its hygroscopicity, namely its capability to adsorb water from the environment, and the surface acquires a hydrophobic character. Zhuravlev predicted the dehydroxylation process to occur at $T_{\text{dh}}= 485\pm 10\unit{K}$. As we show below, our findings agree well with the prediction on the dehydroxylation temperature $T_{\text{dh}}$ for smaller particles ($d= 143\unit{nm}$), but we find larger temperature values for larger particles  ($d= 235\unit{nm}$). We attribute this effect to differences on the NP surface, e.g. surface contaminants or roughness. 


\paragraph*{Mass Loss.}

\begin{figure}
	\includegraphics[width=\linewidth]{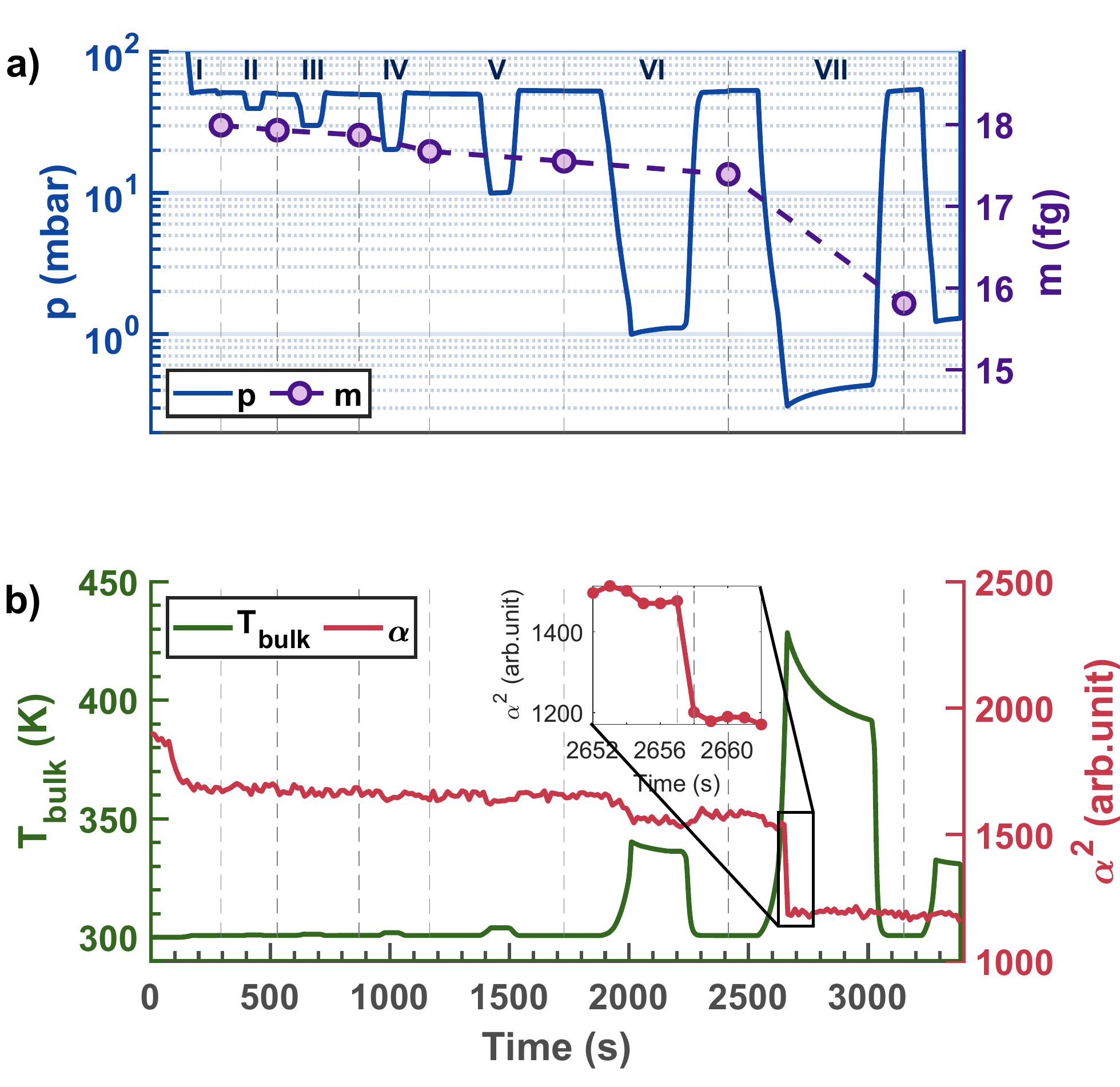}
	\caption{\footnotesize \textbf{Time resolved transition of a levitated NP $(d=235\unit{nm})$ as the pressure is varied.} (\textbf{a}) Pressure $p$ in the vacuum chamber (blue line) and measured mass $m$ (purple circles) over time. Each pressure cycle (I-VII) starts at $p_{s} = 50\unit{mbar}$, where the NP mass is measured ($n_q\neq0$). Then the particle's charge is neutralized ($n_q=0$) and the pressure is cyclically changed. In each cycle, $p$ is reduced further to different final pressures $p_f$. The mass decreases steadily during cycles I-VI, from initially $m = 18\unit{fg}$ down to $m \simeq 17.4\unit{fg}$ at $p \approx 1\unit{mbar}$. For $p < 1\unit{mbar}$ (cycle VII), the NP undergoes an additional and sudden mass loss of $\Delta m \approx 1.6 \unit{fg}$. This mass loss is irreversible. (\textbf{b}) Bulk temperature $T_{\text{bulk}}$ (green) and NP polarizability $\alpha$ (red) are measured over time. The NP's brightness is $\propto|\alpha|^2$, which features a decreasing trend with $p$. We attribute the loss in $\alpha$ to surface water evaporation. $T_{\text{bulk}}\propto p$  (see SI \ref{app:PressTemp}) is estimated directly from the pressure measurement and only deviates significantly from room temperature for $p < 10\unit{mbar}$. The relative decrease of $\alpha$ increases with lower $p_f$, due to increasing $T_{\text{bulk}}$. After cycle VI, $\alpha$ increases again slightly at $p_s = 50\unit{mbar}$ due to water absorption from the background gas. In cycle VII, $T_{\text{bulk}}$ increases above $350\unit{K}$ and the NP experiences a sudden change in $\alpha$ at $t = 2657\unit{s}$, as depicted in the inset. $T_{\text{bulk}}$ spikes at the lowest $p$ to $T_{\text{bulk}}\approx 430 \unit{K}$. Even when $T_{\text{bulk}}$ is reduced back to room temperature again, $\alpha$ stays constant. No additional changes in the properties of the particle are observed after this transformation.}
	\label{fig:02_MassLoss}
\end{figure}

Figure \ref{fig:02_MassLoss} investigates the transformation of a levitated NP as it is brought into vacuum. For this aim, we measure the NP's mass dependence on the bulk temperature $T_{\text{bulk}}$ in a controlled manner by loading a NP in clean air and cycling the pressure between $p_{s} =  50\unit{mbar}$ and decreasing the final pressures to $p_{f} = 50,40,30,20,10,1$ and $0.3\unit{mbar}$ (Fig. \ref{fig:02_MassLoss}a, blue line). During this process, we solely vent the vacuum chamber with N$_2$, such that we have a changing ratio of air and N$_2$ with each cycle. In order to gain additional information about changes in volume and refractive index, we continuously monitor the brightness $\propto\alpha^2$ throughout the whole experiment (Fig. \ref{fig:02_MassLoss}b). To do so, we collect the light scattered by the NP at 90 degrees angle from below the chamber (see SI \ref{app:Brigthness}). The mass is measured always at the same pressure ($p_{s}=50\unit{mbar}$) after each cycle (Fig. \ref{fig:02_MassLoss}a, purple circles) by analysing the NP motion in response to an electric field (with $n_q\neq0$)  \cite{Ricci2019Accurate}. The particle's charge is set back to $n_q=0$ after every mass measurement and before reducing the pressure below $p_{s}=50\unit{mbar}$. Moreover, the NP's bulk temperature $T_{\text{bulk}}$ is also estimated (see Fig.\ref{fig:02_MassLoss}b, green line), as it is proportional to the measured pressure $p$ (see SI\ref{app:PressTemp}). 

The first change that the NP experiences is observed as a moderate reduction of $\alpha$ when the pressure is reduced from atmospheric pressure down to $p_{s}=50\unit{mbar}$ for the first time (cycle I). The next pressure cycles (II-V) produce weak changes in $\alpha$ and a negligible increase in $T_{\text{bulk}}$. The effect of those pressure cycles on the NP is quantified from the mass measurement. As we see in Fig. \ref{fig:02_MassLoss}a, the NP's mass is slightly reduced after each pressure cycle by an overall difference of $\Delta m = 0.6\unit{fg}$ until $p \approx 1\unit{mbar}$. We attribute these changes in $m$ and $\alpha$ to gradual water desorption from the surface, affecting the density $\rho$, the volume ($\propto R^3$), and the refractive index $n_{r}$ of the NP. \\
While $p_{f} > 10\unit{mbar}$, $T_{\text{bulk}}$ is comparable to the room temperature $T_{\text{room}} = 300\unit{K}$ (see Fig. \ref{fig:02_MassLoss}b). However, $T_{\text{bulk}}$ increases with decreasing $p$ once the pressure $p_{f}$ is reduced below $10\unit{mbar}$ (see SI \ref{app:PressTemp}). During cycle VI $(p_f = 1\unit{mbar})$, the NP temperature increases to $T_{\text{bulk}}\approx 350\unit{K}$ and more water is desorbed, which can be seen from the significant drop in $\alpha$ (at $t\simeq 2000\unit{s}$). A similar amount of water is absorbed again from the surrounding gas during the next $p_{s} = 50\unit{mbar}$ interval ($t\simeq 2300\unit{s}$) due to the hydrophilic nature of the NP and the humidity of air. This suggests that the aforementioned changes are due to the reversible process of dehydration and rehydration. 

During the next pump down the dehydration process is completed and the NP reaches a steady state, where no more water is available for desorption (see a small plateau at the beginning of cycle VII around $t=2652\unit{s}$ in the inset of Fig. \ref{fig:02_MassLoss}b). If the pressure is further reduced below $p_{f} < 1\unit{mbar}$, the bulk temperature surpasses a critical value around $T_{\text{bulk}} =400\unit{K}$ and the NP undergoes a rapid and irreversible change. The particle mass drops quickly by  $\Delta m \approx 1.6 \unit{fg}$, more than twice the previous total mass loss. In parallel, we independently observe a sudden reduction of the polarizability by $\Delta\alpha^2\approx 20\%$. Moreover, this transition is accommodated with an abrupt change of the charge number state by $\Delta  n_q = 178$ and the mechanical frequency $\Delta\Omega_x/(2\pi) \approx -  0.75 \unit{kHz}$, a direct measure for the NP's density and refractive index (see SI Fig. \ref{fig:freqdrop}). The inset in Fig. \ref{fig:02_MassLoss}b shows that this drop happens in a fraction of a second $\Delta \tau \ll 1\unit{s}$, suggesting a different mechanism than previous changes. After this event, the polarizability $\alpha$ stays constant, independently of the surrounding pressure. We observe this  transition with all silica NPs investigated. The motional eigenfrequencies of the particle always decrease, indicating a density increase or refractive index reduction or both during the transition. At the same time, the charge change tends towards positive charge states $\Delta n_q > 0$. 

As will be shown later, the slow transformation observed in the trends of $m$ and $\alpha$ prior to cycle VII seems to be due to water layers slowly evaporating from the surface of the NP. As discussed in \cite{Zhuravlev1987,Zhuravlev20110TheSurface}, once the evaporation stopped, a compact layer of silanol groups is left, possibly accommodated with a single, strongly bound water layer. The silanol layer is ejected all of a sudden when $T_{\text{bulk}}$ reaches a critical value, after the entire water reservoir has already evaporated. Once the silianol layer is ejected, $\alpha$ and $m$ are now independent from the surrounding gas pressure, as we can see from Fig. \ref{fig:02_MassLoss}. We attribute this behavior to the fact that, in the absence of the hydrophilic silanol groups on the surface, no water can be absorbed by the NP's surface due to the hydrophobic character of silica \cite{Zhuravlev20110TheSurface}. This transformation has been tested to be irreversible over days.


\paragraph*{Water uptake.}
\begin{figure*}
	\includegraphics[width=1\linewidth]{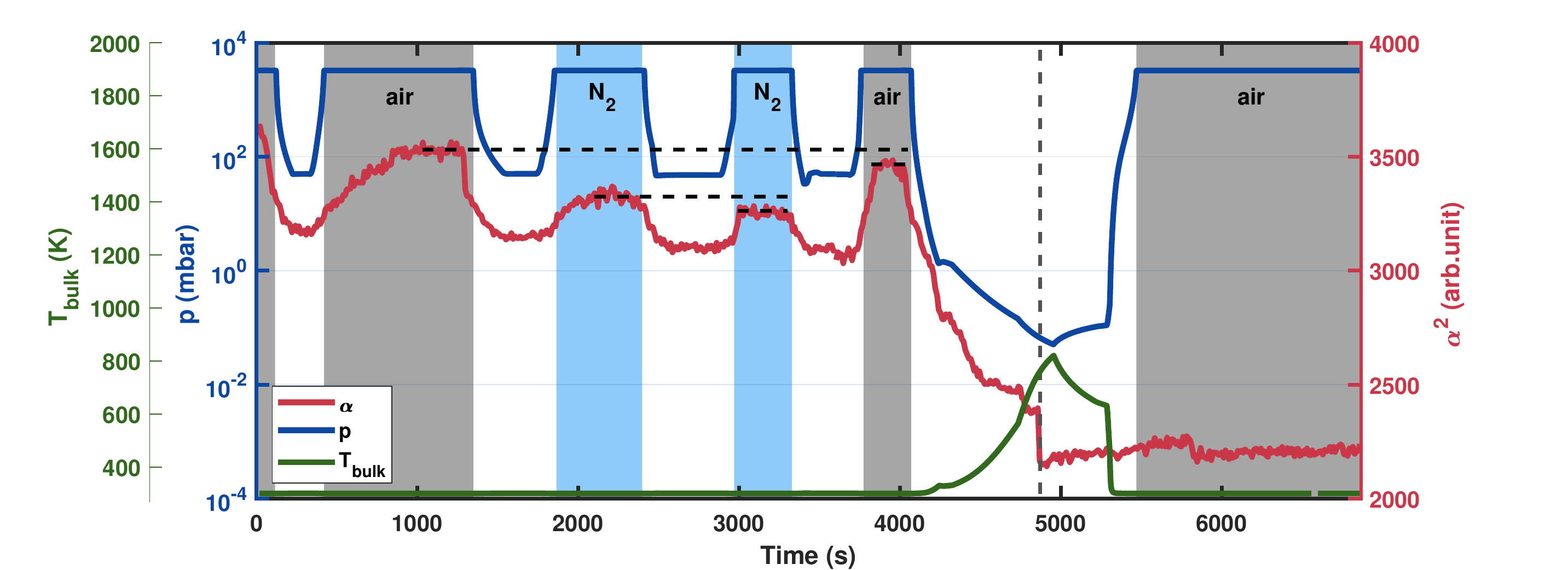}
	\caption{\footnotesize \textbf{Absorption of water by a single NP with $d = 235\unit{nm}$.} The pressure is cyclically varied between atmospheric pressure $p_s$ and $p_{f}= 50\unit{mbar}$, while the gas is changed between nitrogen N$_2$ (blue area) and clean air (gray area). The pressure $p$ (blue), brightness $\propto\alpha^2$ (red), bulk temperature  $T_{\text{bulk}}$ (green) are measured over time. $\alpha$ is following the trend of $p$ and therefore opposing $T_{\text{bulk}}$. This corresponds to water absorption for increasing $p$ and evaporation for decreasing $p$. The first two cycles show that $\alpha$ reaches lower values for the drier N$_2$ than in air, as less water is available in the environment. Comparing two cycles with the same gas, we observe that $\alpha$ fails to reach the initial values, demonstrating only partial reversibility of water absorption. This is observed for both N$_2$ and air. Once $p$ is reduced below  $p_{f}<10^{-2}\unit{mbar}$, corresponding to $T_{\text{bulk}} \approx 750\unit{K}$, the NP undergoes a transition at $t \approx  4850\unit{s}$. Afterwards, $\alpha$ stays relatively constant, meaning that $\alpha$ is $p$ and  $T_{\text{bulk}}$ independent.}
	\label{fig:03_WaterUptake}
\end{figure*}

In order to prove that the variations in polarizability $\alpha$ stem indeed from water evaporation, we exposed a single NP to two background gases with different relative humidity, namely, air and dry nitrogen (N$_2$). Our findings are displayed in Fig. \ref{fig:03_WaterUptake}, where the levitated NP is repeatedly cycled from atmospheric pressure to $p_{f}=50\unit{mbar}$, while changing the type of gas. During this experiment, $\alpha^2$ and $p$ are recorded, while $T_{\text{bulk}}$ is estimated from $p$ (see SI \ref{app:PressTemp}). As it can be seen, our estimates indicate that $T_{\text{bulk}}$ only deviates from room temperature for pressures smaller than $p \approx 10\unit{mbar}$. During the first three cycles with air (once) and nitrogen (twice), we observe that $\alpha$ regains lower values in N$_2$ environments than in air (see horizontal dashed lines). However, if we expose the NP to the more humid air after it was exposed to N$_2$, $\alpha$ increases beyond the value of the previous N$_2$ exposure. We attribute this behaviour to the lower water content in N$_2$ than in air. 
On the other hand, if we compare the values of $\alpha$ for two cycles with the same gas, we see that $\alpha$ fails to completely recover, as also observed in Fig. \ref{fig:02_MassLoss}b. Therefore, the process is only partially reversible on short timescales. The relative difference in $\alpha$ between the first and the second cycle with the same gas is comparable for both air and N$_2$. We hypothesize that reaching higher surface water content requires higher gas humidity, keeping in mind that the NPs were initially dispersed in water prior to being nebulized and subsequently trapped.

If the pressure is reduced below the critical value $p_{\text{crit}} \approx 0.1\unit{mbar}$, corresponding to $T_{\text{bulk}} \approx 750K$, the NP undergoes the aforementioned transition (at $t \approx 4850\unit{s}$), indicated by a discontinuity in $\alpha$. Once the NP has gone through the  transition ($t > 4850\unit{s}$), $\alpha$ stays well below its pre-drop value independently of the pressure. We attribute the following slight oscillation in  $\alpha$ ($5000 < t <5800\unit{s}$) to the rapid venting process and related mechanical stress on the experimental apparatus. Remarkably, $\alpha$ has been reduced by a total of approximately 40\% from the beginning of the experiment. The attained state where the NP is insensitive to ambient humidity is due to the fact that the surface is now hydrophobic, as opposed to the initially hydrophilic character. 

\paragraph*{Size effects.} We studied the size dependence of $T_{\text{bulk}}$ using two different NP sizes, namely $d=143\unit{nm}$ and $235\unit{nm}$. We expect that the critical $T_{\text{bulk}}$ is reached at different pressures $p_{\text{crit}}$ for different sizes, since the laser absorption scales with the size $d$, while the cooling rate of the NP by background gas collisions scales as $\propto d^2$ (see SI \ref{app:PressTemp}). The critical pressure $p_{\text{crit}}$ and hence $T_{\text{bulk}}$ of the  transition is statistically distributed as shown in the SI Fig. \ref{figApp:pressurestats} and \ref{figApp:ptats}. The distribution of critical temperatures at which the  transition takes place are $T_{\text{bulk}} = 550\unit{K}$ and $T_{\text{bulk}} = 860\unit{K}$ for $d= 143\unit{nm}$ and $d=235\unit{nm}$, respectively. 
Notice that, in both cases, the widths of the distributions are $\simeq 200\unit{K}$. We also estimated the uncertainty in the calculation of the critical temperature, which stems from uncertainties in the particle size, pressure in the chamber, and trapping power. The estimated relative uncertainties are always smaller than $17\%$ (see SI \ref{app:DropStats}).
\\
We attribute the unexpected differences in $T_{\text{bulk}}$ for various sizes and broad distributions to pressure uncertainties, NP size distribution and possibly surface composition (e.g. surface roughness or contaminants). Future experiments will be more accurate and therefore more conclusive by deducing the pressure directly from the power spectral density of the particle motion, whose width is directly linked to the pressure.


\paragraph*{Discussion and Conclusions.}
Our platform enabled the observation of the dehydration and dehydroxylation of a single silica NP in real time. 
We gained detailed insights on the time scale of the changes affecting mass, density, refractive index, and surface charge that are inaccessible with other techniques. 
The versatility of our nano-reactor with its unprecedented time, charge, mass and spatial resolution paves the way to new insights in surface chemistry of single NPs, preventing ensemble effects.  
The technique with its high level of control can readily be extended to simultaneously studying several NPs by using multiple optical tweezers, and to other types of chemical processes using different background gases.   
Other materials, e.g. more absorbing ones, could also be investigated with non-optical trapping mechanisms \cite{Conangla2018Motion} or counter propagating beams \cite{Donato,divitt2015cancellation}, readily available in the field of levitodynamics. Moreover, the ability to generate centrifugal forces in a controlled way by particle rotation \cite{kuhn2017optically,Reimann2018,ahn2018optically, monteiro2018optical} gives access to the investigation of binding forces. Combined with the control of the bulk temperature in the range of $100K < T_{\text{bulk}} < 1000$K, it also opens the door towards hot and cold chemistry, respectively. 
Finally, the evaluation of the hygroscopicity of aerosols, i.e., their ability to absorb water from the surrounding gas, is a particularly relevant and active topic, since the water content alters significantly the size of aerosol particles, which critically affects the climate \cite{Kaufman2002, Ramanathan2119,cotterell2014measurements, valenzuela2020optical, valenzuela2020testing,diveky2021shining}. We believe that our demonstrated capability to measure the hygroscopicity of silica can contribute to the study of aerosols in the future.\\


\textbf{Acknowledgments}.\hspace{0.2cm} The authors thank L.\:Novotny for stimulating discussions. The project acknowledges financial support from the European Research Council through grant QnanoMECA (CoG - 64790), Fundaci\'{o} Privada Cellex, CERCA Programme / Generalitat de Catalunya, and the Spanish Ministry of Economy and Competitiveness through the Severo Ochoa Programme for Centres of Excellence in R$\&$D (SEV-2015-0522). 
RAR acknowledges financial support from FEDER/Junta de Andaluc\'ia-Consejer\'ia de Transformaci\'on Econ\'omica, Industria, Conocimiento y Universidades/Projects C-FQM-410-UGR18, P18-FR-3583, and A-FQM-644-UGR20. AWS acknowledges funding through the Deutsche Forschungsgemeinschaft (DFG, German Research Foundation) under Germany’s Excellence Strategy – EXC-2123 QuantumFrontiers – 390837967.
\\

\pagebreak


%
\cleardoublepage


\pagebreak
\newpage
\appendix

\subsection{Production of aerosol}\label{app:aerosol}
We produce microsized droplets of an ethanol-water-nanoparticle suspension in the aerosol phase with the help of a medical nebulizer. We worked with NPs in the range $d  = 100-300$nm. In order to avoid the creation of clusters, we mix a solution of $3\unit{mL}$ ethanol with 2.5~$\mu$L nanoparticle-water-suspension (5 weight percent) where each droplet has a probability $\rho\lesssim \:1$ to contain one NP. This suspension is nebulized through a metal funnel to direct the NP flow towards the trapping region and possibly slowing the droplets down to increase the trapping probability. We achieve a trapping event around every second of which $50\%$ can be used for experiments. Cluster trapping events can be identified by measuring the brightness of the trapped object in real time.

\subsection{Experimental Setup}\label{app:ExpSetup}
\begin{figure}[h]
\includegraphics[width=\linewidth]{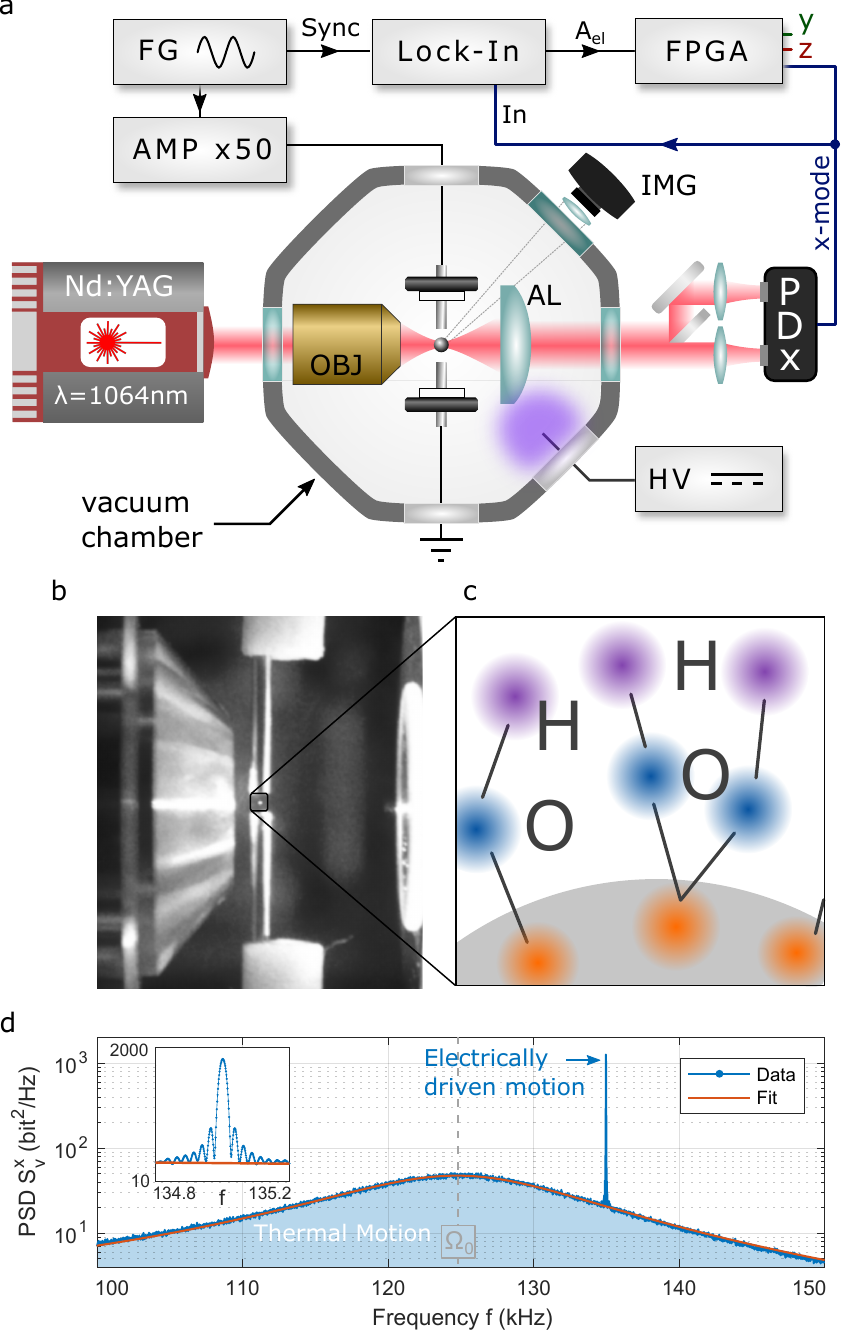}
\caption{\footnotesize \textbf{Experimental setup.} (\textbf{a}) Schematic depiction of the setup. (\textbf{b}) A real picture of the setup inside of the vacuum chamber. (\textbf{c}) Illustration of silanol groups on the particle surface.  (\textbf{d}) PSD of the thermally and electrically driven motion of the NP at its natural mechanical frequency $\Omega_0$ at $p=50\unit{mbar}$. The inset shows in detail the particle's response to electronic drive used for charge and energy determination.}
\label{figApp:01_Exp_SetUp}
\end{figure}

A pair of electrodes are connected to the amplified signal generated by a function generator (FG), creating an electric field that  drives motion of the charged NP. The purple glow on the side of the chamber is emitted by a plasma generated by a bare electrode connected to a high voltage (HV) DC source, and used to control the net charge of the NP \citep{Frimmer2017Controlling}. 

The experimental setup is depicted in SI Fig. \ref{figApp:01_Exp_SetUp}a. A $\lambda = 1064\unit{nm}$ laser (power $P \in [50 - 100] \unit{mW}$) is focused by an objective (OBJ, NA = 0.8) and traps a single silica NP in the focus. The light scattered by the NP in the forward direction and the part of the trapping beam that do not interact with the nanoparticle are collected with an aspheric lens (AL) and detected with a split detection scheme (PDX) in order to infer the motion of the NP from the interference pattern of both beams. An FPGA and a lock-in amplifier are used to bandpass and record the signal from the detector. The signal thus obtained is proportional to the NP`s motion that describes a thermally and harmonically driven, damped resonator. Two electrodes form a parallel plate capacitor (see SI Fig. \ref{figApp:01_Exp_SetUp}b), that creates a sinusoidal electric field from the voltage signal generated by a signal generator and amplified with a high voltage amplifier (AMP). The electric field acts on a charged particle via the Coulomb force, enabling charge and mass measurements.
In SI Fig. \ref{app:ExpSetup}c, an illustration of the silanol groups on the particle surface is depicted. \\ 
SI Fig. \ref{app:ExpSetup}d shows the thermally driven power spectral density (PSD) of the particle motion, centered at $\Omega_0/2\pi$. The additional peak corresponds to the electrically driven motion (see inset). The peak height depends on the particle charge $n_q$ and the particle mass $m$.
\subsection{Charge control of a levitated nanoparticle}\label{app:ChargeControl}

\begin{figure}
	\includegraphics[width=0.48\textwidth]{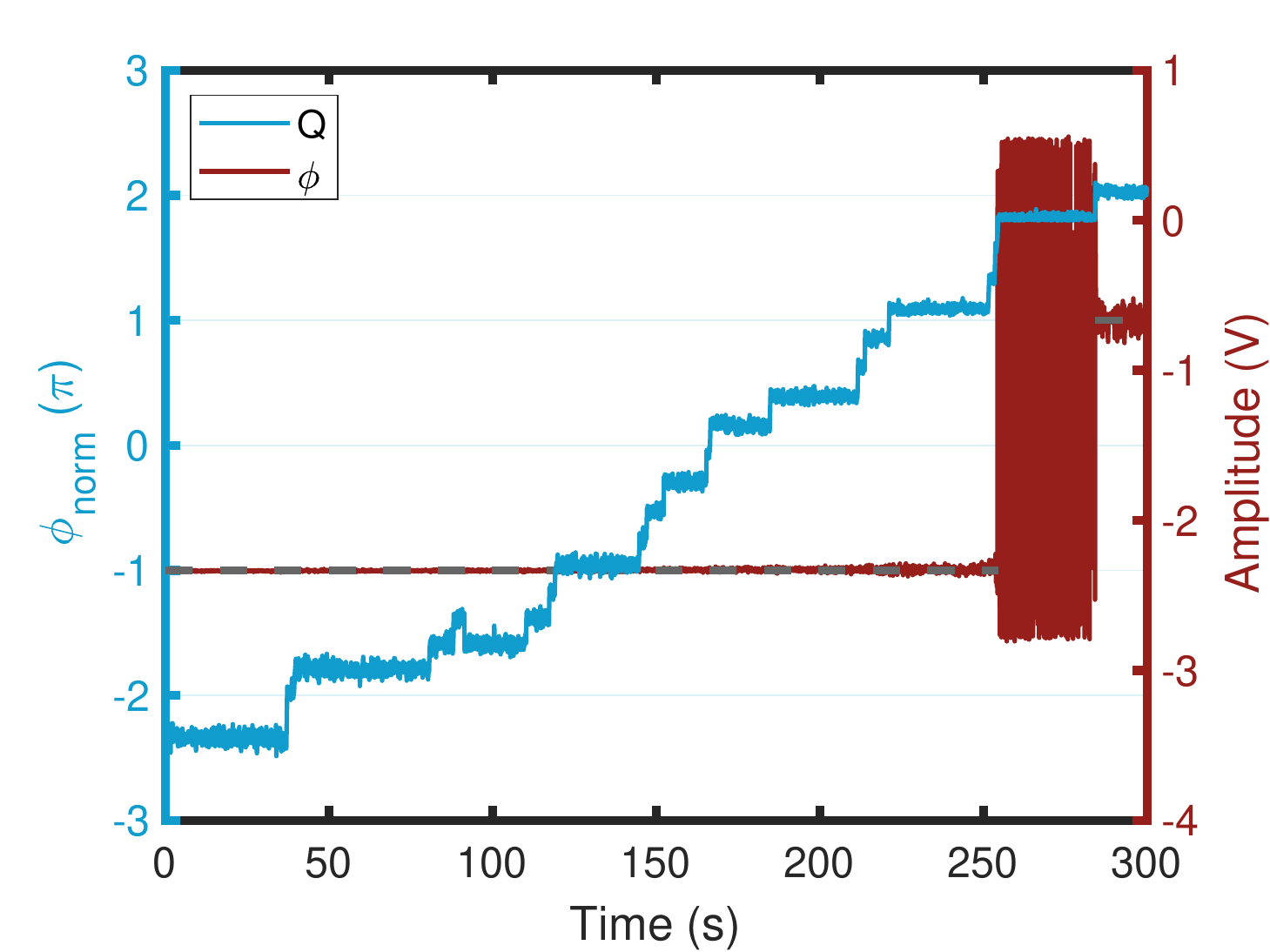}
	\caption{\footnotesize \textbf{Charge control of NP:} The particle's surface charge $n_q$ (blue) is manipulated in well defined single elementary charge steps. The phase of the driven particle motion (red) is well defined until the particle is uncharged at $n_q =0$. }
	\label{fig:chargesteps}
\end{figure}

The number of charges $n_q$ is controlled by applying a high DC voltage $U_{dc} \approx \pm 1\unit{kV}$  on a bare electrode placed on one side of the vacuum chamber, creating a corona discharge that gives rise to a plasma consisting of a mixture of ions (polarity depending on $U_{dc}$ polarity) \citep{Fridman2004Plasma}. Positive and negative ions are accelerated toward opposite directions due to the presence of the electric field from the electrode. As a result, the ratio of positive to negative charges reaching the particle is biased by the electrode polarity, thus allowing us to fully control the final charge of the particle within positive or negative values on the single elementary charge level as depicted in SI Fig. \ref{fig:chargesteps}. The amplitude of the driven peak (blue) shows equidistant steps, corresponding to individual charges. The phase between the driving signal and the NP motion (red) is fixed, until the NP is uncharged ($n_q =0$) and no force drives the NP. 

\subsection{Mass measurement}\label{app:Mass}
The detailed mass measurement procedure can be found in \cite{Ricci2019Accurate}. Here, we give a short summary of the procedure. An oscillating Coulomb force  $\pmb {F_{el}}$ drives a charged NP ($n_q = 1-10$) at $\omega_{dr}$ and produces a peak in the PSD of amplitude $S_v(\omega_{dr})$. The solely electric contribution is given by $S_v^{el}(\omega_{dr})= S_v(\omega_{dr}) - S_v^{th}(\omega_{dr})$, where $S_v^{th}(\omega_{dr})$ is the undriven thermal response. The ratio $R_s = S_v^{el}(\omega_{dr})/S_v^{th}(\omega_{dr})$ allows one to calculate the particle mass $m$ as $$m = \frac{n_q^2 q_e^2 E_0^2 \tau}{8 k_B T \Gamma R_s} ,$$ where $\Gamma$ is the oscillator's mechanical damping and $\tau$ the measurement duration. We perform this procedure at $p = 50\unit{mbar}$ to avoid contributions of the trap nonlinearities to the measurement.


\subsection{Brightness \& Rayleigh Scattering}\label{app:Brigthness}

Consider a plane wave scattered by a spherical particle of radius $R$ and refractive index $n_r$. The plane wave is characterized by the wave vector $\textbf{k}$ and electric field \textbf{E}. The dipole moment $\textbf{p}=\alpha \textbf{E}$ of the particle in air or vacuum is calculated from the polarizability of the particle $\alpha$, which for a uniform sphere takes the form:
\begin{equation}
	\alpha =4\pi\varepsilon_0 R^3\frac{n_r^2-1}{{n_r^2+2}}
\end{equation}
The polarizability is related to the scattering cross section for Rayleigh particles ($R\ll \lambda/10$) 
\begin{equation}
	\sigma_{\rm scat}=\frac{k_0^4}{6\pi}|\alpha|^2,
\end{equation}
where $k_0$ is the wavenumber in vacuum. The brightness of the scattered light that is measured by the camera is directly proportional to $\sigma_{\rm scat}$. Therefore one can monitor changes in both $R$ and/or $n_r$ from the variations of the detected brightness. \\
For our case of the small particle size $d= 143nm$, the Rayleigh approximation is valid. The larger particle size of $d=235nm$ falls into the transition regime between Rayleigh and Mie scattering. Nevertheless, this does not pose an issue, since we are only interested in relative brightness. Furthermore, the Rayleigh scattering cross section for the larger particles used here differs only by a factor of two in between the Rayleigh approximation and the Mie scattering theory \cite{cox2002experiment}.


\subsection{Pressure and Bulk Temperature}\label{app:PressTemp}
The steady state $T_{\text{bulk}}$ of an isolated NP is given by the equilibrium between absorbed and emitted power from and to the environment. The heating of a levitated NP are due to absorbed laser power and black-body (BB) absorption, while the cooling happens through residual gas collisions and BB emission. Under our experimental conditions, the main heating source is the absorbed laser power $P_{\text{abs}}$, even in cases of low absorbing materials as silica. The main cooling source is heat exchange between the NP and the residual gas molecules $P_{\text{gas}}$. At high vacuum, BB radiation is the dominating cooling source $P_{\text{BB emit}}$, and for completeness we also consider the absorption due to BB $P_{\text{BB abs}}$. The thermal equilibrium is governed by the power balance: 

\begin{equation}
    P_{\text{abs}} + P_{\text{BB abs}} = P_{\text{BB emit}} + P_{\text{gas}}
    \label{eq:powerbalance}
\end{equation}

\paragraph*{Heating by absorbed laser power.}
A NP of volume $V  = \frac{4}{3} \pi R^3$ trapped in a focused laser beam with a waist $w_t = \frac{\lambda}{\pi NA}$ and intensity at the centre of the trap  $I_0 = 2 P/(\pi w_t^2)$ absorbs \cite{Chang2010Cavity}
  
$$P_{\text{abs}} = 12 \pi \frac{I_0 V}{\lambda} \Im\left(\frac{\epsilon -1 }{\epsilon + 2}\right),$$
where $\epsilon = \epsilon' - j \epsilon'' = n_r^2 = n'^2 + k^2 -2 j n'k $.  We assume the complex refractive index to equal $n_r = n' - j k = 1.46 + j\: 5.98 \times 10^{-8}$ (corresponding to losses of  3dB/km). 
Hence, 

$$\Im\left(\frac{\epsilon -1 }{\epsilon + 2}\right) =  -\frac{ 3\epsilon''}{(\epsilon'+2)^2 + \epsilon''^2}$$

\paragraph*{Cooling by residual gas.}
The background gas consists mainly of nitrogen molecules N$_2$ with a molecular mass $m_{gas}= 2\times 14\:au$ and $au = 1.66053904\times 10^{-27}kg$. The molecules move with a speed of $v_{rms} = \sqrt{3k_B T_{\text{room}}/m_{gas}}$ where $T_{\text{room}}$ is room temperature and $k_\text{B}$ the Boltzmann constant. The gas specific heat ratio is $\gamma_{sh} = 7/5$ for an ideal gas. The power, by which the NP is cooled, is then given as

$$ P_{\text{gas}} = \alpha_g \sqrt{\frac{2\pi}{3}} \frac{d}{4}^2 p \: v_{rms} \frac{\gamma_{sh}+1}{\gamma_{sh}-1} \left(\frac{T_{\text{bulk}}}{T_{\text{room}}}-1\right)$$

The phenomenological energy accommodation factor is $\alpha_g = 0.61$ \cite{Hebestreit2018Measuring,Hebestreit2017Thermal}.

\paragraph*{Black-body radiation (Heating \& Cooling).}
The power absorbed and emitted by BB radiation at temperature $T$ is given by
$$P_{\text{BB}}(T) = \frac{dE}{dt} = \frac{72 \zeta(5) V}{\pi ^2 c^3 \hbar^4} (k_B T)^5 \Im\left(\frac{\epsilon-1}{\epsilon+2}\right)$$

with the Riemann zeta function $\zeta(5)\approx 1.04$. The absorbed and emitted power equals $P_{\text{BB abs}}(T_{room})$ and $P_{\text{BB emit}}(T_\text{bulk})$, respectively.\\

$T_{\text{bulk}}$ is then given as the temperature that fulfills Eq. \ref{eq:powerbalance}, and is obtained from numerical solution with standard routines.

\subsection{ Transition statistics}\label{app:DropStats}

\begin{figure}
	\includegraphics[width=0.48\textwidth]{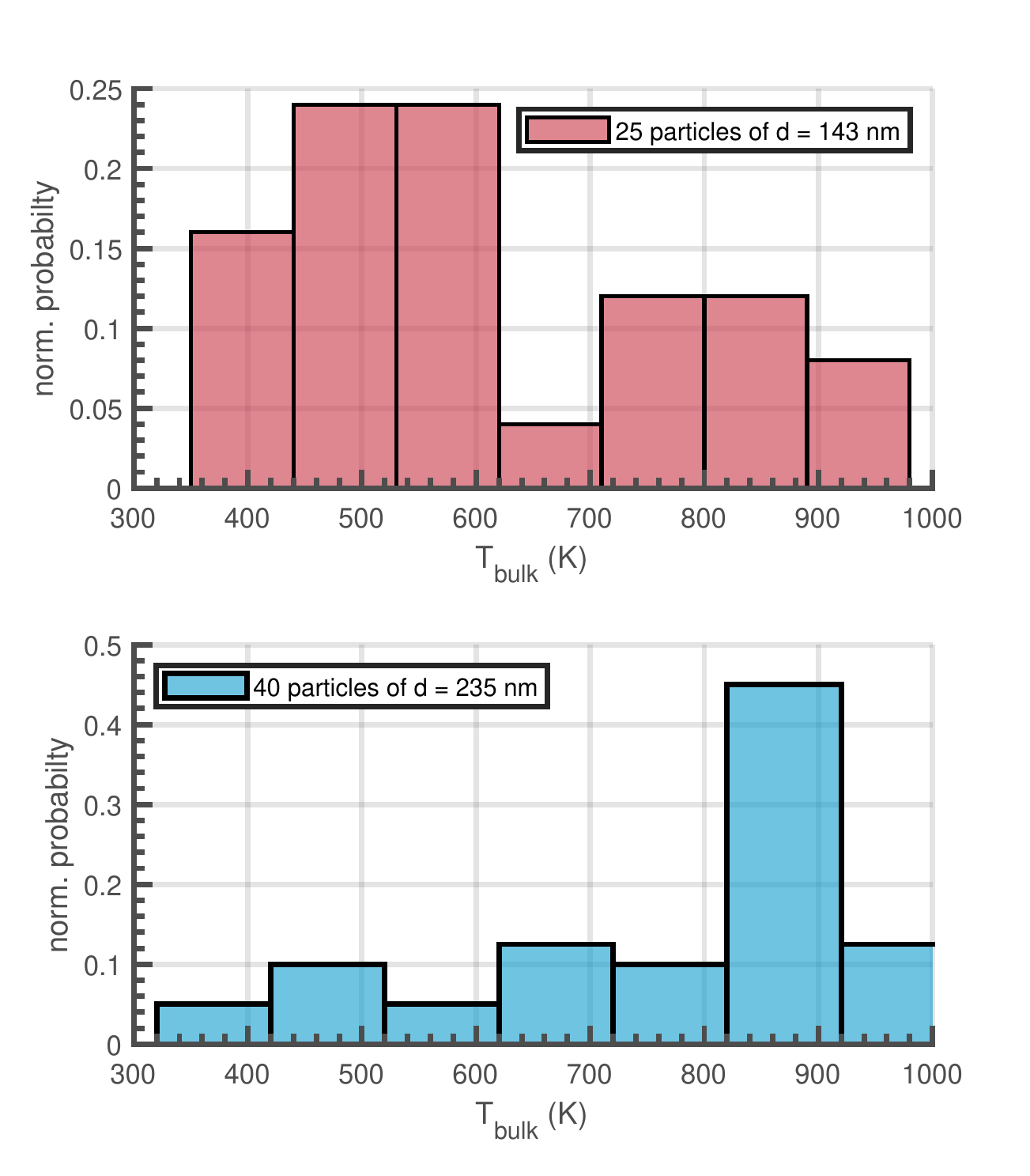}
	\caption{\footnotesize \textbf{Normalized probability distribution of the critical bulk temperature $T_{\text{bulk}}$. Top:} for $d = 143\unit{nm}$ with a maximum at $T_{\text{bulk}}\approx 550\unit{K}$ of $\approx 25\%$. The sample size is 25 NPs. \textbf{Bottom:} for $d = 235\unit{nm}$ with a maximum at $T_{\text{bulk}}\approx 860\unit{K}$ of $\approx 45\%$. The sample size is 40 NPs.}  
	\label{figApp:pressurestats}
\end{figure}

\begin{figure}
	\includegraphics[width=0.48\textwidth]{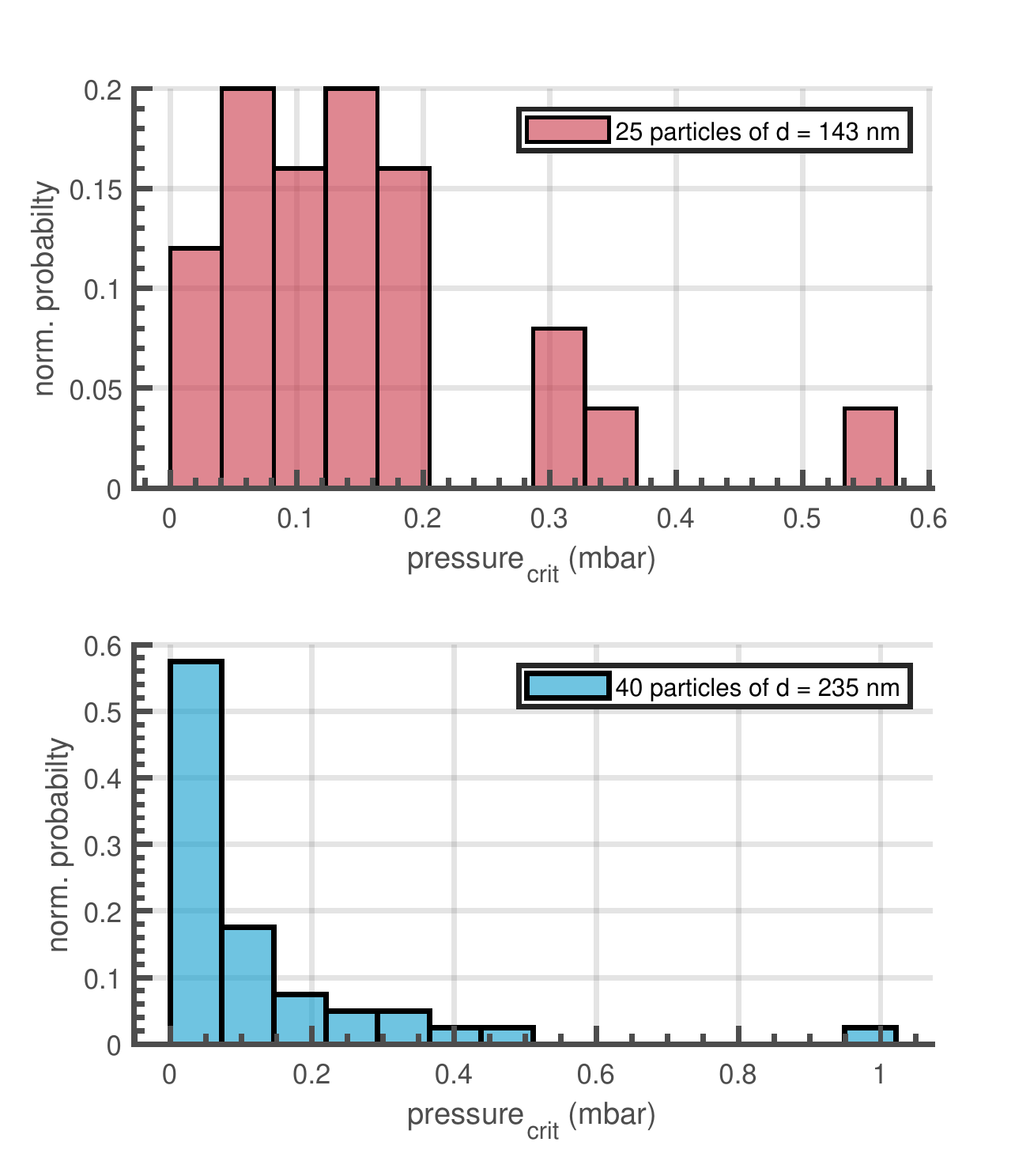}
	\caption{\footnotesize \textbf{Normalized probability distribution of the critical pressure $p_{\text{crit}}$. Top:} for $d = 143\unit{nm}$ with a maximum at $p_{\text{crit}}\approx 0.1\unit{mbar}$ of $\approx 15-20\%$. The sample size is 25 NPs. \textbf{Bottom:} for $d = 235\unit{nm}$ with a maximum at $p_{\text{crit}}\approx 0.04\unit{mbar}$ of $\approx 60\%$. The sample size is 40 NPs.}  
	\label{figApp:ptats}
\end{figure}

In the present study, two types of NPs with different diameters are investigated: $d = 143 \pm 4 \unit{nm}$ and $d = 235 \pm 11 \unit{nm}$ (nominal value of the manufacturer \cite{microparticles2018Data}). For each particle, we estimate $T_{\text{bulk}}$ at which the abrupt  transition occurs. Its normalized probability distribution is depicted in Fig. \ref{figApp:pressurestats}. For a sample of 25 particles of $d=143\unit{nm}$, the most probable $T_{\text{bulk}}$ is $\approx 550\unit{K}$. The distribution shows a double peak structure with a second peak at $T_{\text{bulk}}\approx 850\unit{K}$. In the case of $d=235\unit{nm}$ with a sample size of 40 particles, the most probable $T_{\text{bulk}}$ equals $\approx 900\unit{K}$. The distribution is asymmetric towards lower temperatures but shows a clear maximum. For completeness, Fig. \ref{figApp:ptats} depicts the normalised probability distribution of the critical pressure $p_{\text{crit}}$.

\begin{figure}
	\includegraphics[width=0.4\textwidth]{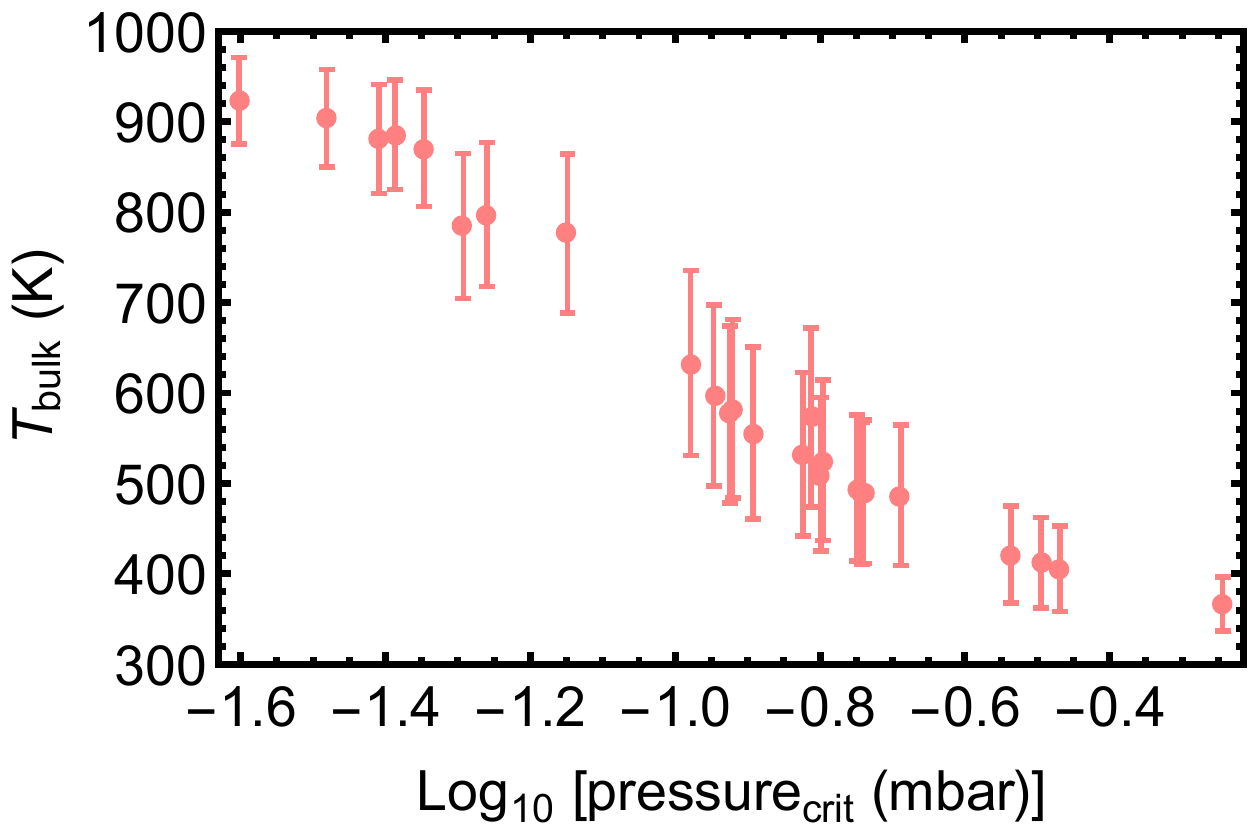}
	\includegraphics[width=0.4\textwidth]{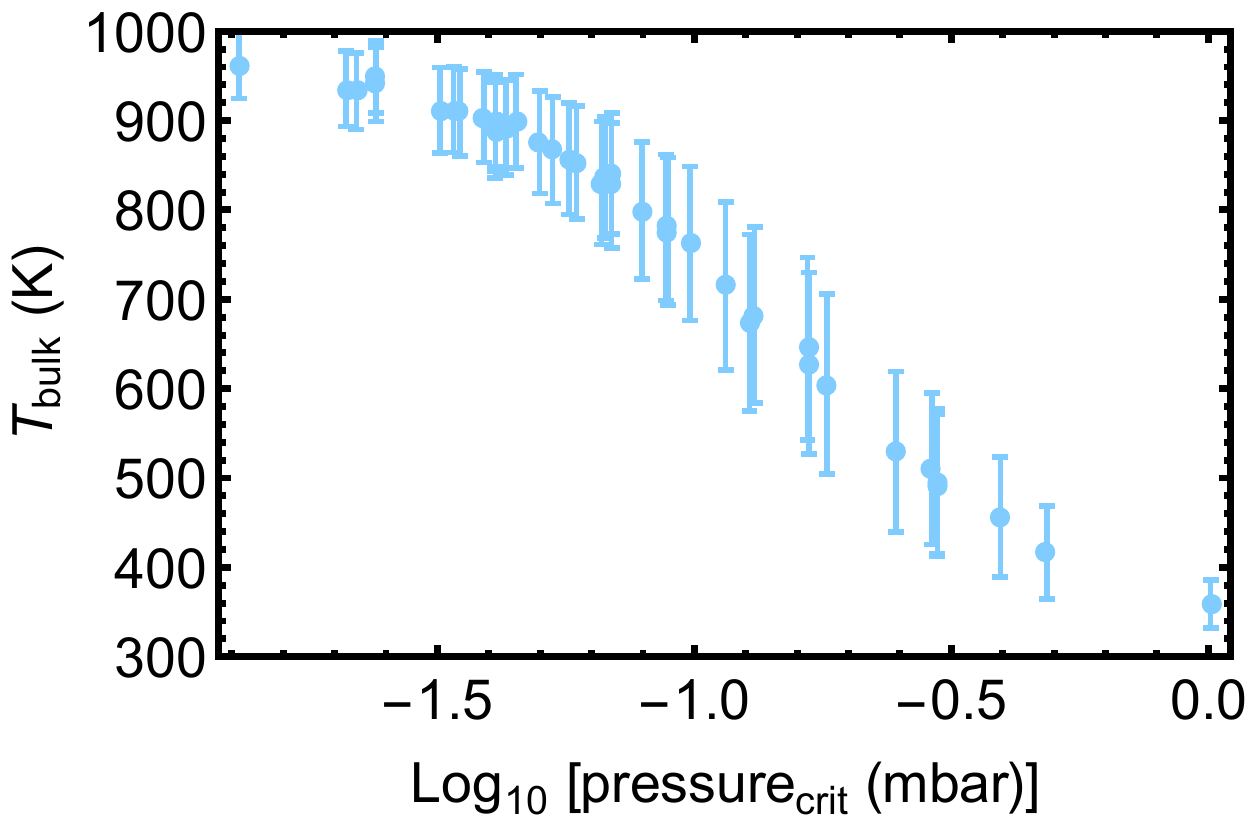}
	\caption{\footnotesize \textbf{Calculated critical bulk temperature $T_{\text{bulk}}$ and its uncertainties versus pressure. Top:} for $d = 143\unit{nm}$.  \textbf{Bottom:} for $d = 235\unit{nm}$.}  
	\label{figApp:Tbulk_pressure}
\end{figure}

\begin{figure}
	\includegraphics[width=0.4\textwidth]{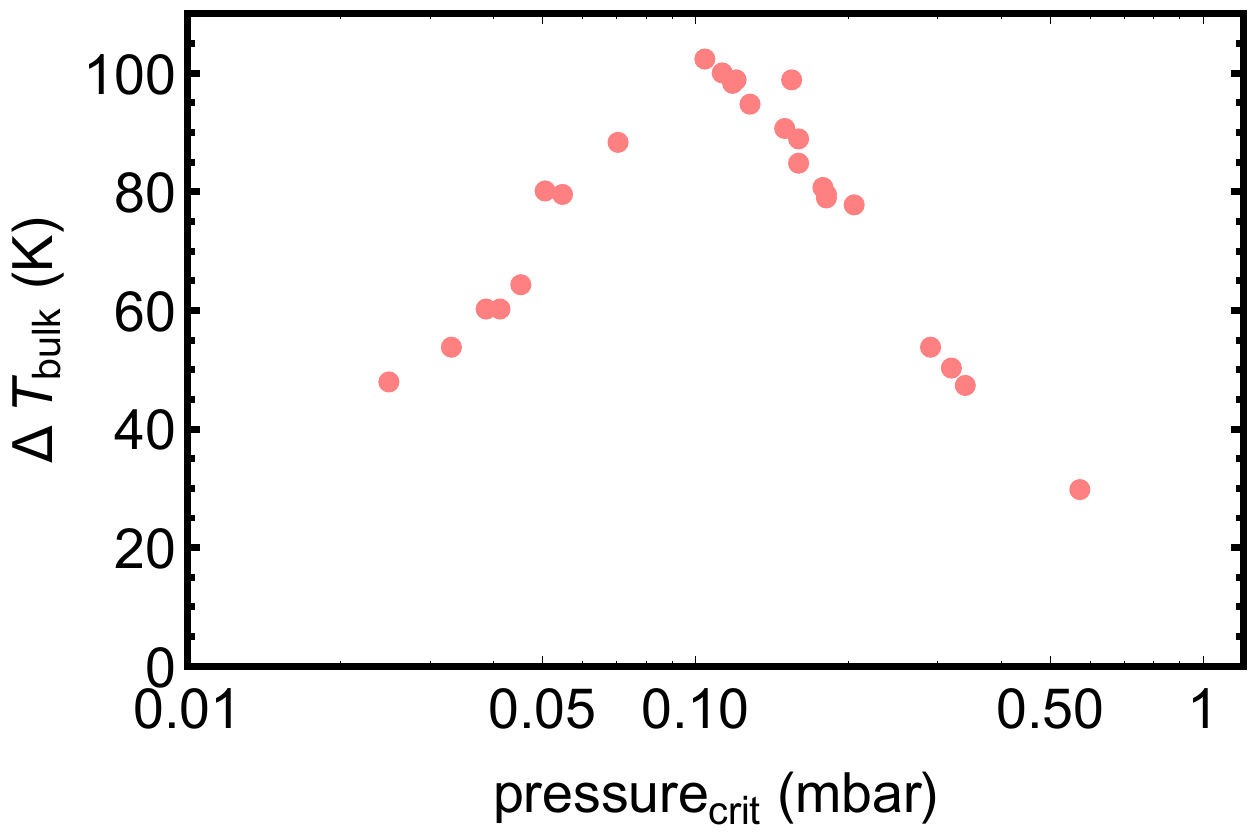}
	\includegraphics[width=0.4\textwidth]{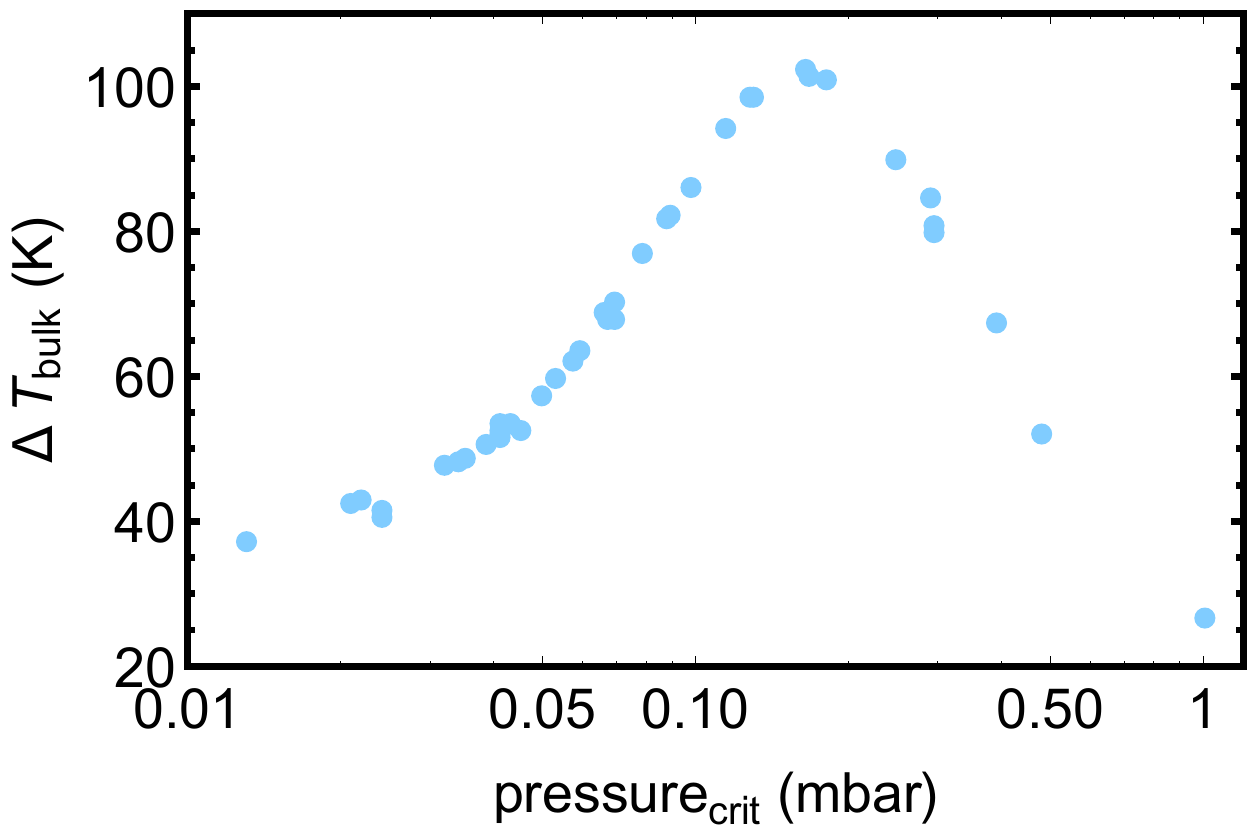}
	\caption{\footnotesize \textbf{Uncertainties of calculated critical bulk temperature $T_{\text{bulk}}$ versus critical pressure. Top:} for $d = 143\unit{nm}$.  \textbf{Bottom:} for $d = 235\unit{nm}$.}  
	\label{figApp:errors}
\end{figure}

The obtained values of the critical temperature are accompanied by uncertainties due to different effects. First, the particle size as given by the manufacturer is estimated to be within $3\%$ of the nominal value. Moreover, the accuracy of the pressure gauge can have up to $30\%$ of uncertainty, and an uncertainty in the trapping power of $1\unit{mW}$ was also considered. 
The uncertainties $\Delta T_{\text{bulk}}$ in the calculation of the critical temperature were estimated by varying individually the input parameters (optical power, pressure, particle size) in Eq. \ref{eq:powerbalance} according to their individual uncertainties, e.g. $T_{\text{bulk}}(d \pm \Delta d)$. The individual temperature uncertainty is estimated as the maximum of $\Delta T_{bulk}(d) =  |T_{bulk}(d)- T_{bulk}(d\pm\Delta d)|$ where  $T_{\text{bulk}}(d)$ is obtained from Fig. \ref{figApp:pressurestats}. The total temperature uncertainty is then estimated as $ \Delta T_{bulk} =\sqrt{\Delta T_{bulk}(d)^2 + \Delta T_{bulk}(p)^2 + \Delta T_{bulk}(P)^2} $. The results are shown in Fig. \ref{figApp:Tbulk_pressure} and \ref{figApp:errors}. 
As it can be seen, this method provides uncertainties up to $\sim 100\unit{K}$, which appears at $T_{\text{bulk}}\simeq650\unit{K}$ for both particle sizes. Interestingly, the position of this maximum uncertainty  coincides with the pressure value at which both terms from gas collisions and black body radiation contribute. The uncertainties appear smaller in the two regimes where only one mechanism dominates, like gas collisions for $T_{\text{bulk}}\ll 650\unit{K}$ and black body radiation for  $T_{\text{bulk}}\gg 650\unit{K}$. In each of these regimes there is only one source of uncertainty, Hence, reducing the total uncertainty.\\
The relative uncertainties are always below $17\%$, and do not seem to compromise the validity of our conclusions.

\begin{figure}
	\includegraphics[width=0.48\textwidth]{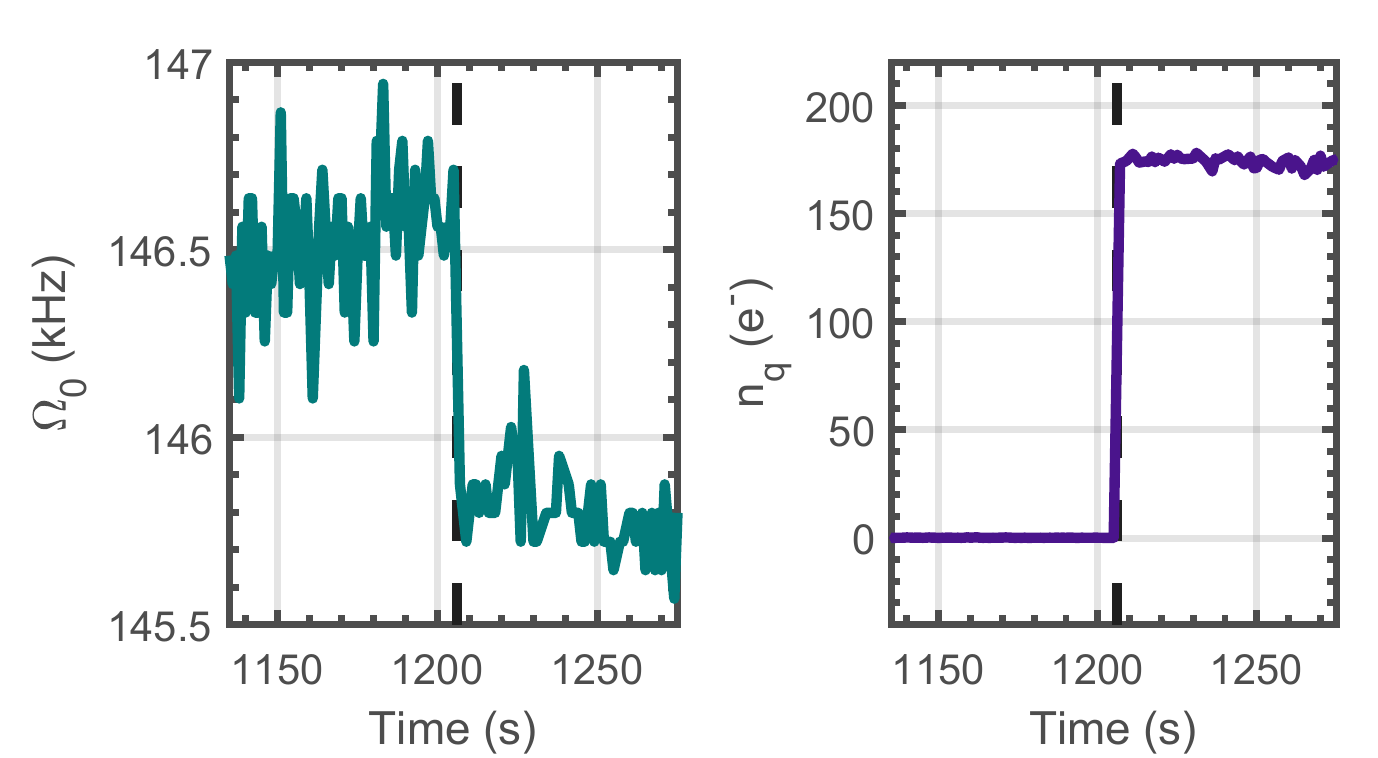}
	\caption{\footnotesize  \textbf{Time resolved example of the  transition of a NP}. Left panel: eigenfrequency $\Omega_0/2\pi = \Omega_x/2\pi$ jumps from initially $146.5\unit{kHz}$ to $145.75\unit{kHz}$, corresponding to $\Delta \Omega_x/2\pi = 0.75\unit{kHz}$. This indicates a density increase or refractive index reduction of the NP. Right panel: number of elementary charges $n_q$ gained during the  transition yield 178$e^-$.}
	\label{fig:freqdrop}
\end{figure}

\newpage


%

\end{document}